\documentclass{article}
\usepackage[T1]{fontenc}
\usepackage[utf8]{inputenc}
\usepackage{ismir} 
\usepackage{amsmath,cite,url}
\usepackage{graphicx}
\usepackage{color}

\title{Muse-it: A Tool for Analyzing Music Discourse on Reddit}

\oneauthor
  {Jatin Agarwala \hspace{1cm} George Paul \hspace{1cm} Nemani Harsha Vardhan \hspace{1cm} Vinoo Alluri}
  {International Institute of Information Technology, Hyderabad\\\texttt{jatin.a@research.iiit.ac.in}}

\begin{document}

\maketitle

\begin{abstract}
Music engagement encompasses a spectrum of interactions between individuals and music—ranging from the selection of music and the emotional responses it evokes to its influence on behavior, identity, and social connections. Social media platforms provide forums where users engage organically in online conversations on a variety of topics, offering a setting in which music behavior can be observed as it naturally occurs. Advances in natural language processing (NLP) and big data analytics now enable researchers to leverage these discussions for large-scale analysis and extend music research to broader contexts. Reddit is one such platform that provides anonymity, allowing diverse users to participate in dialogues without inhibition. These Reddit threads offer researchers rich discourse and real-world opinions on musical associations and engagement in an ecological setting. However, the nature of big data necessitates advanced tools to extract, store, process, visualize, and analyze it. In this paper, we introduce Muse-it, a platform designed to retrieve comprehensive data and create datasets centered on specific queries. It enables researchers to retrieve data from all subreddits that feature posts related to a specific query. In addition, it performs topic modeling, temporal trend analysis, and clustering of posts and comments, helping researchers study large amounts of data efficiently. Muse-it also identifies hyperlinks to music streaming platforms like Spotify and collects corresponding musical features of the tracks, providing metadata such as track and artist names, album name, release date, genre, popularity, and lyrics, as available. Complementing this functionality is an interface that generates dynamic visualizations of the data. In sum, Muse-it is an easy-to-use tool that enables music researchers to collect big data related to music and gain insights into music behavior as it naturally occurs.

\end{abstract}

\begin{figure}[h]
    \centering
    \includegraphics[width=1\linewidth]{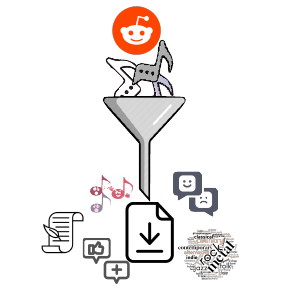}
    \caption{We introduce Muse-it, a tool to retrieve,  process, visualize, and analyze music-related Reddit discourse using search queries. Furthermore, it provides metadata of the tracks, albums, and playlists linked in the extracted data for further analysis.}
    \label{fig:Visual}
\end{figure}

\section{Introduction}\label{sec:introduction}

Reddit is a social media platform built around user-created communities called subreddits \footnote{\url{https://www.reddit.com/best/communities/1/}} where users can interact anonymously. While other social media platforms have certain limitations due to users having a social desirability bias or the website having character limits, Reddit provides users with complete anonymity which encourages them to share their experiences freely and openly on a wide range of topics. This unique feature of Reddit makes it an excellent source for collecting diverse and organic user experiences. Another reason why Reddit is a desirable platform for music research is due to the fact that people often post links to their personal music choices, including playlists, in discussions or as recommendations to others. This acts as a more organic way to retrieve data on musical experiences and to examine the relationship people have with music. 

A range of studies has demonstrated the versatility of Reddit data across different subfields within MIR, and music perception and cognition, challenging or confirming existing literature. For example, \cite{if_i_like_2024} scraped approximately 5000 threads from the subreddit r/ifyoulikeblank where users request and provide music recommendations. They compared human recommendations with algorithmic recommendations, highlighting their differences - with human recommendations also taking extra-musical factors into considerdation. \cite{asmr_article} analyzed Reddit comments to delve deeper into the relation between ASMR and music-induced frisson, and found that ASMR is consistent with Huron's theory of frisson \cite{huron_2006}, that negatively valenced responses can be supressed to result in an overall positively valenced effect.

The anonymity that Reddit affords its users allows vulnerable populations to discuss sensitive topics openly. For example, \cite{bhavyajeet} analyzed posts from mental health subreddits (e.g., r/depression and r/depressionMusic) to identify adaptive and maladaptive music-listening strategies. They found that maladaptive music-listening is associated with certain acoustic features and lyrical properties, a crucial finding for designing recommendation systems to regulate mood and related disorders. Similarly, \cite{sharon2024} examined discussions in r/autism and used various statistical methods to identify musical associations and preferences of individuals with Autism Spectrum Disorder (ASD). While a lot of their findings align with previous research, the popularity of pop and electronica among individuals with ASD is a finding that contradicts previous observations and highlights variability within the ASD spectrum. These studies highlight Reddit’s ability to capture nuanced, first-person accounts of the psychological impacts of music, data that is not easily accessible via traditional surveys or lab experiments due to their sensitive nature.

The presence of subreddits allow researchers to explore niche communities. Studies such as \cite{moisio2022just} and \cite{mishra-etal-2021-metal} have explored specific subreddits, examining the identity construction among fans of specific music genres and the emotional associations with them. However, the vast amount of data present on Reddit enables broader research as well. \cite{Veselovsky_Waller_Anderson_2021} created a novel dataset of 1.3M instances of music sharing on Reddit and presented a set of methodologies to understand the macro-scale structure of online music sharing and its relations with various social and cultural dimensions.

Despite the promising insights offered by these diverse applications, MIR research has yet to fully exploit the potential of Reddit data. The technical complexity of compiling and processing large-scale datasets from Reddit and the technical knowledge required to organize all this information in a manner that would allow researchers to extract meaningful inferences poses a significant barrier, particularly for researchers whose expertise lies outside of computational methods.

\begin{figure}[h]
    \centering
    \includegraphics[width=1\linewidth]{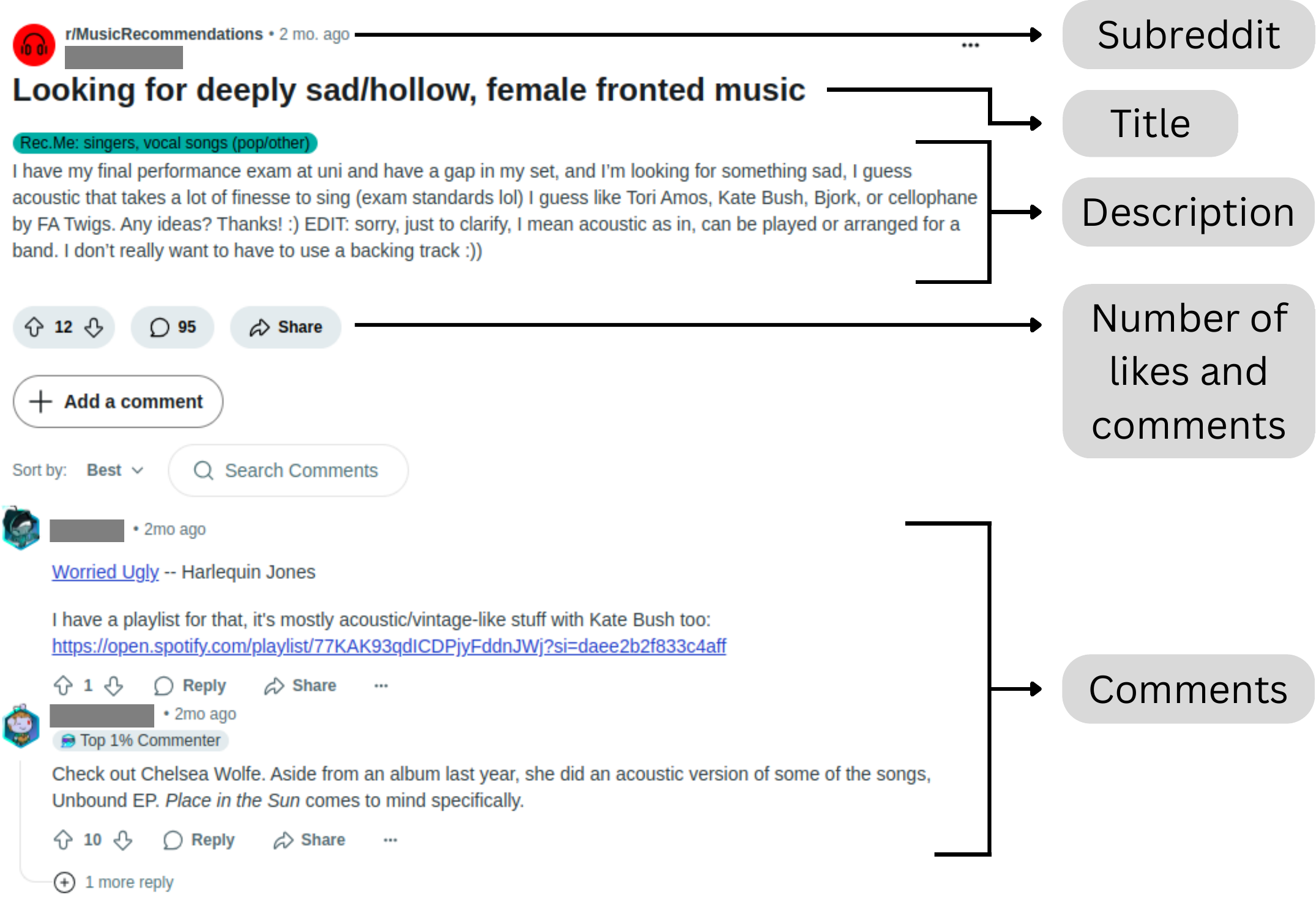}
    \caption{Structure of a Reddit Post}
    \label{fig:reddit_post}
\end{figure}

In this paper, we introduce \textit{Muse-it}\footnote{Muse-it is available at \url{https://github.com/Muse-it/MuseIt} for Windows, Linux, and macOS. However, there are a few differences required in the configuration of the tool according to platform. These differences, along with the installation and setup instructions, are available in the README of our online code repository at \url{https://github.com/Muse-it/MuseIt/blob/master/README.md}.}, a tool designed to streamline the retrieval, analysis, and visualization of Reddit data for music research. Muse-it automates the extraction of posts and comments across relevant subreddits, generating comprehensive datasets centered on custom queries. The platform incorporates advanced analytical techniques such as topic modeling, temporal trend analysis, hierarchical clustering of topics, among others, to reveal patterns in music-related discourse. Moreover, Muse-it identifies hyperlinks to music streaming platforms like Spotify \footnote{\url{https://open.spotify.com/}} and uses SpotDL\footnote{\url{https://spotdl.readthedocs.io/en/latest/}} to fetch detailed metadata of the tracks, playlists, and albums identified. This includes track names, artist names, album details, release dates, genres, popularity metrics, and lyrics, among others. We believe that Muse-it can enable music cognition researchers to take advantage of the vast amount of content present on social media platforms such as Reddit in an efficient and easy-to-use manner, to streamline investigations into music’s societal, emotional, and cultural roles at scale. By lowering technical barriers, the tool supports interdisciplinary exploration of music’s diverse impacts, empowering researchers to generate novel hypotheses and validate existing findings in real-world settings like Reddit.

Muse-it offers a powerful means of retrieving and analyzing big data. It can expedite traditional mixed-methods research by making big data more accessible and manageable for researchers. The detailed datasets generated by Muse-it can provide critical insights and help formulate hypotheses, which can then be further examined using controlled, lab-based studies. In this way, the tool serves as an aid for researchers, lowering the technical barrier to harnessing big data in the study of music behavior and cognition. 

In the following sections, we describe the underlying architecture and functionality of Muse-it. We use results for the search query "sad music" to generate the figures in this paper.\footnote{A demo of the tool along with the data used to create the visualizations can be found at \url{https://drive.google.com/drive/folders/1DtQZ7oRwrp4LPSu8-EajPTWX_l93WOs4?usp=sharing}.}

\begin{figure*}
    \centering
    \includegraphics[width=1\linewidth]{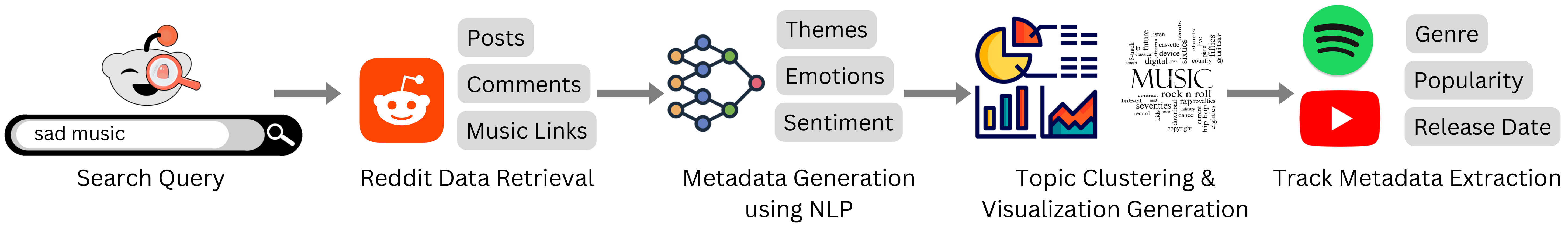}
    \caption{Muse-it pipeline}
    \label{fig:pipeline}
\end{figure*}

\section{Muse-it architecture and functionality}

Muse-it comprises four modules as depicted in Figure \ref{fig:pipeline}. The first module comprises Data retrieval from Reddit. The second module aids in metadata extraction like emotion, sentiment and topic from the text. The third module allows the researcher to create visualizations of the metadata and performs hierarchical clustering of the topics. The final module uses the Spotify links found in the collected data and extracts the metadata of the tracks. These modules are explained in detail in the following subsections.

\begin{figure}[h]
    \centering
    \includegraphics[width=1\linewidth]{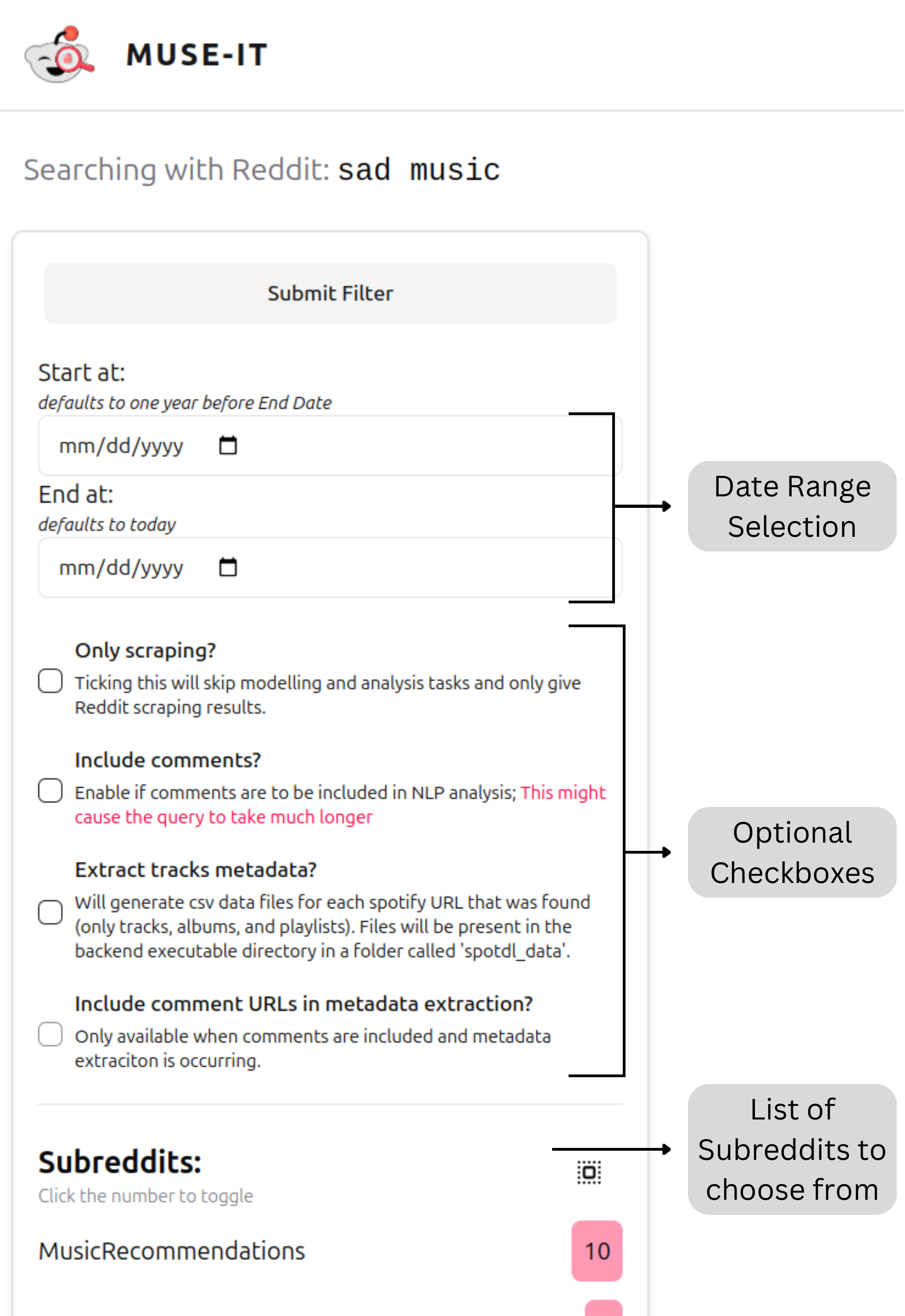}
    \caption{Muse-it interface for selecting filters for retrieving Reddit data for the search query. Note that the fourth checkbox to include comment URLs in track metadata extraction can only be enabled if both the options before it are enabled.}
    \label{fig:filter_screen}
\end{figure}

\begin{figure}[h]
    \centering
    \includegraphics[width=1\linewidth]{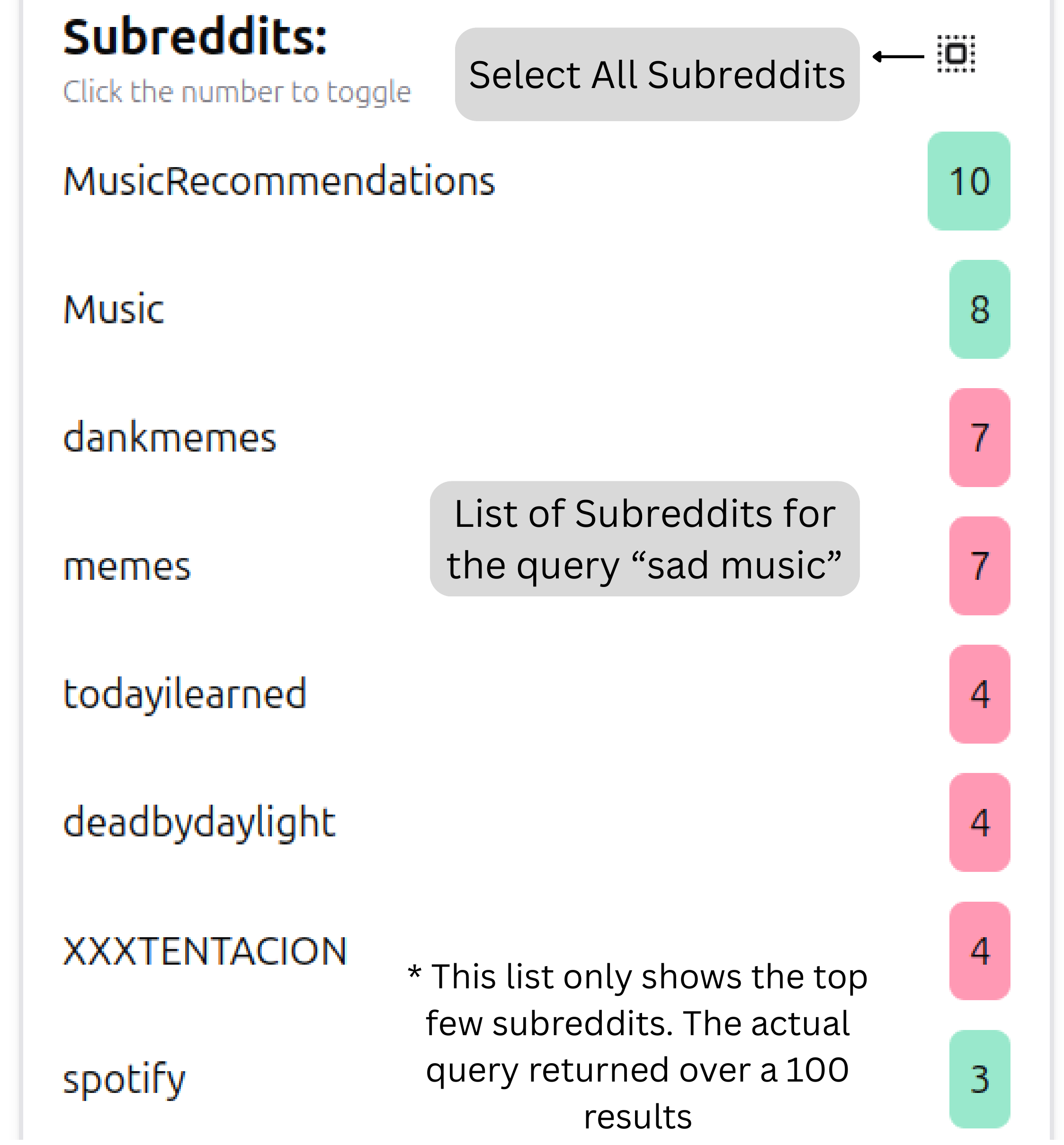}
    \caption{List of subreddits for the query "sad music". A subreddit can be included for the retrieval by clicking on the number next to it, which turns it green from red. The numbers beside the subreddit show the number of posts found in the subreddit matching the query in a quick preliminary search by the Reddit API. While these numbers can be insightful, they tend to be biased towards subreddits with more frequent activity.}
    \label{fig:subreddits_list}
\end{figure}

\begin{figure*}
    \centering
    \includegraphics[width=1\linewidth]{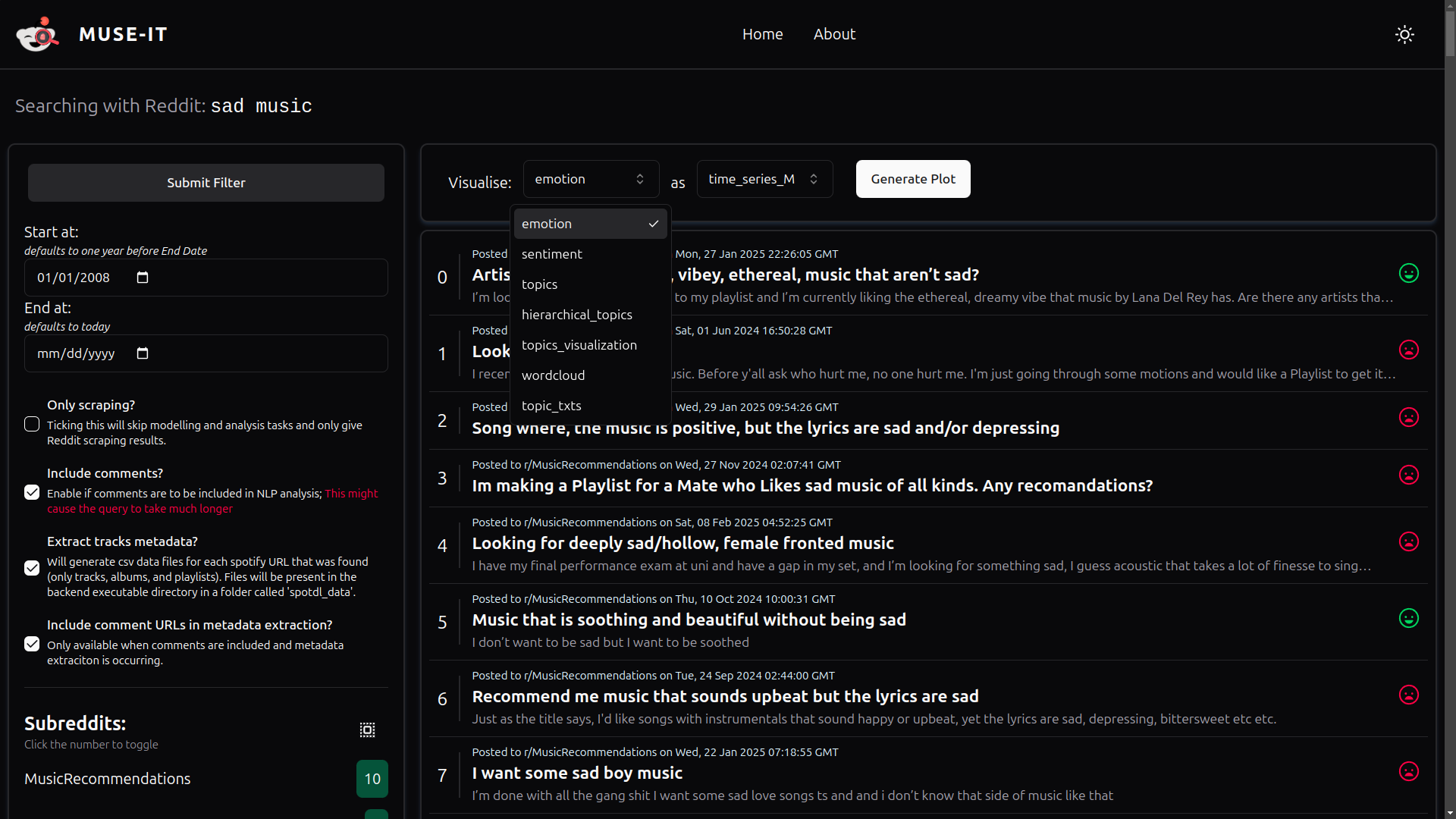}
    \caption{This figure shows the Muse-it interface after retrieval is completed for the query "sad music". Muse-it allows researchers to visualize various features using a drop down menu and download them once the information retrieval is complete. It also displays the posts retrieved along with other metadata such as subreddit and time of creation. In this figure, the interface has been switched to the "Dark mode" using the button on the top right of the screen.}
    \label{fig:download_screen}
\end{figure*}

\subsection{Reddit Data Retrieval}
Data retrieval is conducted using the Reddit API\footnote{\url{https://developers.reddit.com/docs/api}} and Python Reddit API Wrapper (PRAW)\footnote{\url{https://praw.readthedocs.io/en/stable/}}. The initial step is utilizing the specified query to search for a list of subreddits that reference or discuss the query topic. The next step involves selecting the subreddits to be searched. The same query is used to retrieve results from all selected subreddits. For example, Figure \ref{fig:subreddits_list} shows the list of subreddits for the query "sad music". The ones selected are turned green in color. Note that one limitation of the Reddit API is that it allows for the extraction of only 1000 posts per subreddit. Muse-it also provides various filters to cater to each researcher's unique requirements, as shown in Figure \ref{fig:filter_screen}. These filters are discussed in greater detail in the relevant sections.

The Reddit API provides the name of the subreddit, post title, post body, post URL, number of comments, and time of creation. The tool searches the query over the selected subreddits using Reddit's search API and retrieves a large number of relevant results. The tool also processes all of the links mentioned in the post and filters for Spotify URLs, which is used in the fourth module to extract track metadata. All this is done in approximately 5-10 minutes for every 10,000 posts, depending on the size of the post. Muse-it also provides an option to retrieve the comments under each post. However, this can slow down the process as the number of comments increases. If comments are retrieved, Muse-it also creates a separate column for comment URLs, which contain the Spotify URLs mentioned in the comments of the post. 

\subsection{Metadata Generation using NLP}
\begin{figure}[h]
    \centering
    \includegraphics[width=1\linewidth]{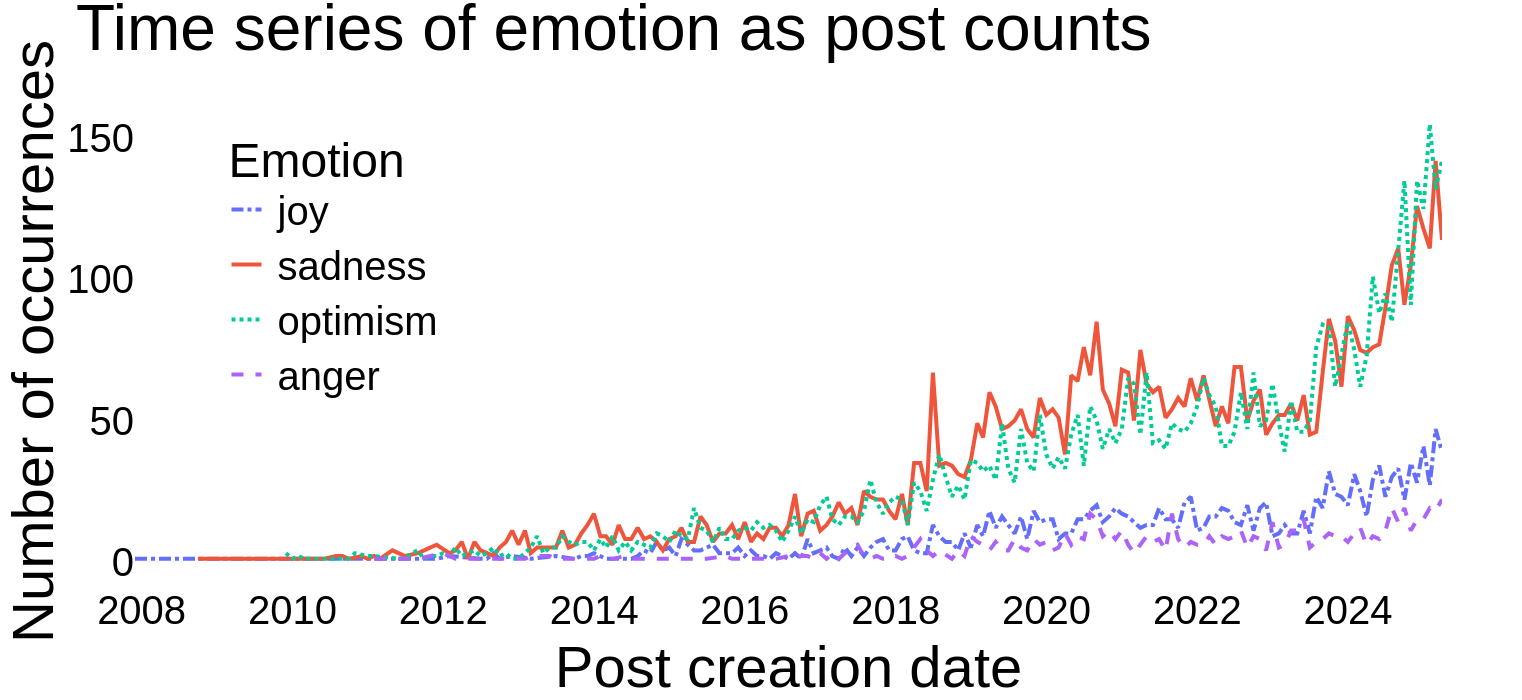}
    \caption{Monthly time-series distribution of emotions in the posts retrieved for the query "sad music". Similar plots are generated for daily and weekly temporal trends. Interestingly, we see a distinct drop in optimism and increase in sadness in the year 2020 around the start of the global Covid-19 pandemic.}
    \label{fig:emotion_ts}
\end{figure}

Muse-it processes the retrieved data to generate secondary metadata, which includes themes, emotions, and sentiments from the post titles. Figure \ref{fig:emotion_ts} shows the temporal trends in emotion for the query "sad music". We primarily use titles since they convey the general opinion and mood of the post and are more computationally efficient versus longer texts. Since the generation of this secondary metadata can be a slow process and might not be required in all use cases, it is provided as an optional feature that can be disabled via a checkbox in the interface.

Muse-it uses the TweetNLP \cite{camacho-collados-etal-2022-tweetnlp} library and their open-source models for text-classification of themes, emotions, and sentiments. All of the models are transformer-based language models trained on Twitter data and then fine-tuned for task-specific classification. TweetNLP takes about 4 minutes per 10,000 posts. Despite being trained on Twitter data, which is known for being concise, the creators of the model found that it performs well on larger contexts as well, including Reddit \cite{antypas2023robust}, which is why it was chosen. Including other models within Muse-it and giving researchers the choice between them remains a future possibility.


\subsection{Topic Clustering and Visualization Generation}

Muse-it generates visualizations for all the extracted data and metadata. It generates interactive and dynamic visualizations, such as topic modeling and clustering, and shows the distributions of each metadata attribute, including themes, emotions, and sentiments. It also plots the changes in these distributions over time to show the trends of these attributes. 

\begin{figure}[h]
    \centering
    \includegraphics[width=1\linewidth]{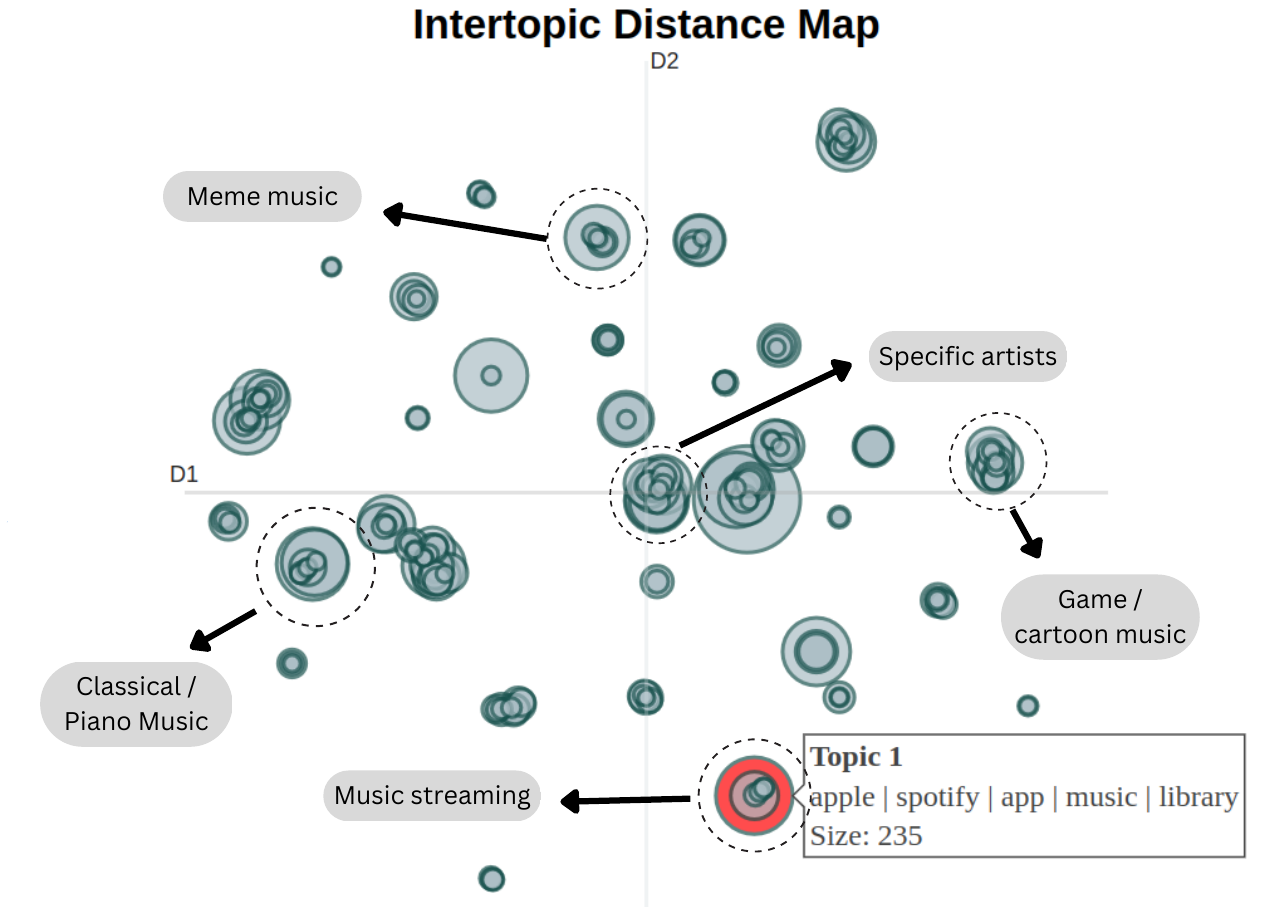}
    \caption{Dynamic visualization of the topics as clusters in a 2D plane. We have identified and labeled some clusters with dotted circles. The interface also provides a slider to highlight individual topics, which has been used to highlight Topic 1 in this figure.}
    \label{fig:clustering_visualization}
\end{figure}

Muse-it uses BERTopic\cite{grootendorst2022bertopic} to create a hierarchical clusters of the topics found in the posts. It uses Sentence BERT\cite{reimers-2020-multilingual-sentence-bert} to generate embeddings for the entries and then uses those embeddings to visualize the topics and their hierarchy. Muse-it allows researchers to visualise the heirarchical clustering in the form of a dendogram. As shown in Figure \ref{fig:clustering_visualization}, the platform also creates a dynamic visualization of the same on a 2-dimensional plane that clearly shows the distances between the topics. We have highlighted and labeled some prominent clusters in the figure, like classical/piano music on the bottom left which contains topics such as "soft piano music", "sad piano", "classical piano music", etc.

\begin{figure}[h]
    \centering
    \includegraphics[width=1\linewidth]{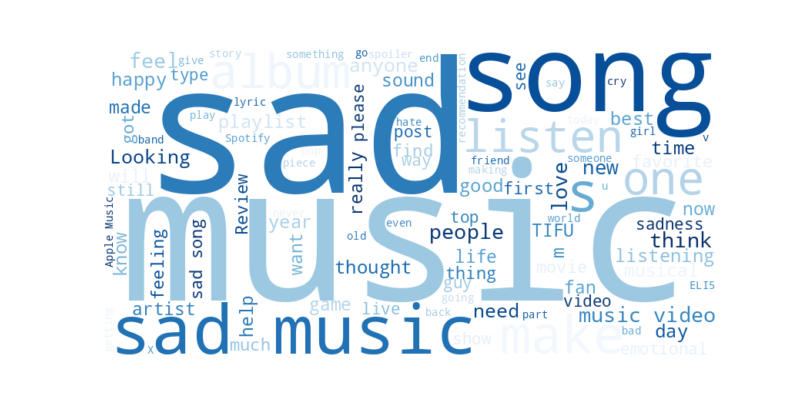}
    \caption{Wordcloud generated for posts on "sad music".}
    \label{fig:wordcloud}
\end{figure}

Muse-it also generates word clouds, which are good at showing the frequency and relevance of words within a text corpus. It helps in gaining an understanding of the themes in the retrieved data, allowing the researcher to gather an overview of the data in a quick and efficient manner.

\subsection{Track Metadata Extraction}

Muse-it processes links mentioned in the posts and identifies the Spotify URLs. If the Spotify URL belongs to a track, album, or playlist, we process it further by using SpotDL to collect details such as \textit{track name, artist name, album name, genres, date, year of release, lyrics, duration, publisher, copyright text, }and whether it is an explicit song. This is done for each track in the case of a URL that points to an album or a playlist. By default, this is only done for the URLs in the body of the post. Muse-it provides functionality (via an optional checkbox) to retrieve track metadata for the URLs present in the comments of the post as shown in Figure \ref{fig:filter_screen}. This is not a default setting as there can be a large number of URLs present in the comments for popular posts.

On average, the metadata of each track in a playlist or album with 100 tracks takes about 2 minutes. We identified that some playlists shared on Reddit can have over 1000 songs, which can significantly slow down the extraction process. Therefore, we recommend a timeout of 5 minutes, or approximately 250 songs. If a URL takes longer than the timeout duration to process, it is terminated and Muse-it moves on to the next URL. This timeout duration, by default, is set to the recommended 5 minutes and is a parameter that can be changed in the configuration file.

\subsection{Downloading the Data}

\begin{figure}[h]
    \begin{center}
        \includegraphics[width=0.7\linewidth]{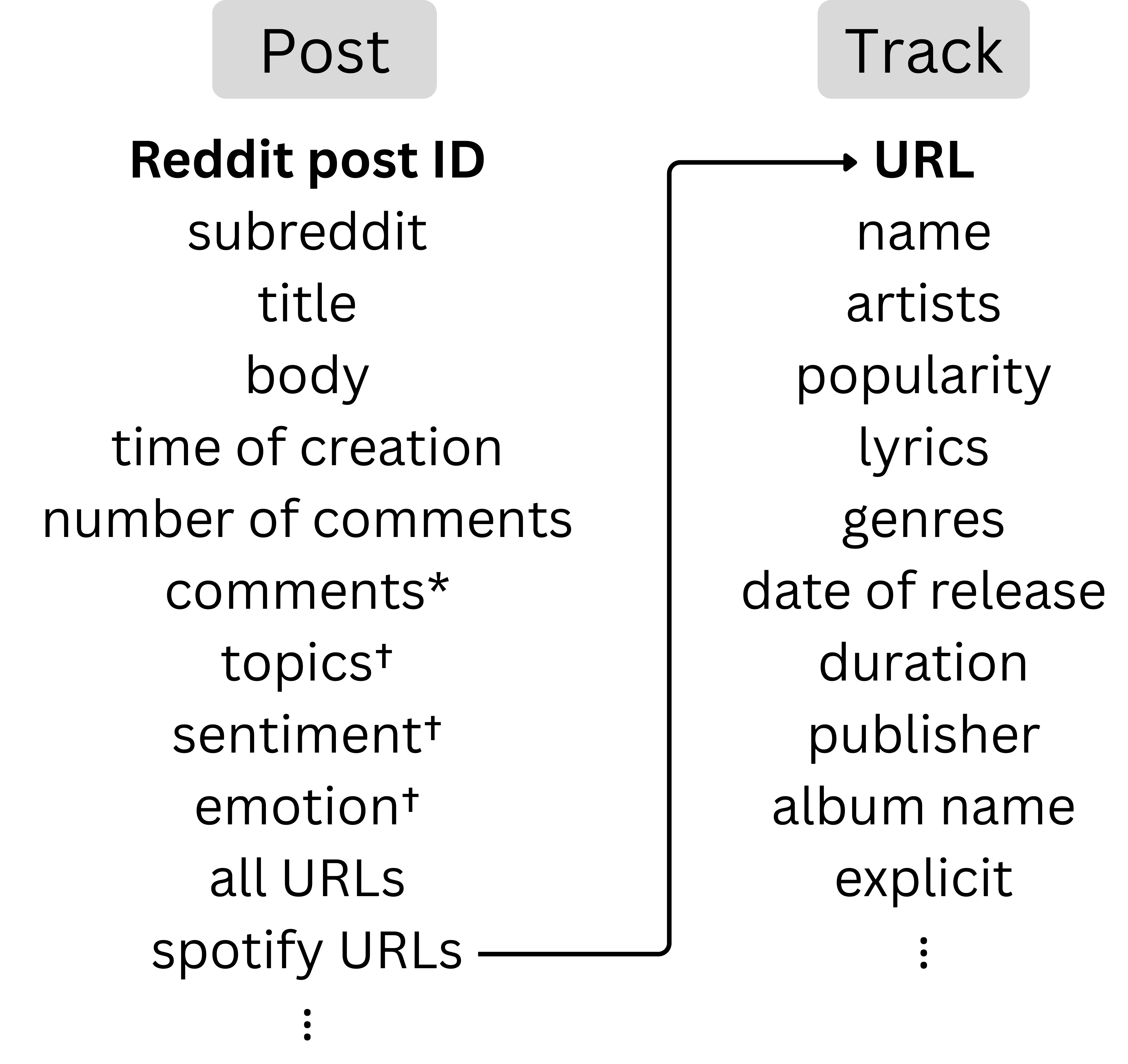}  
    \end{center}

    * : Comments are present only if the "Include comments?" checkbox is selected.
    
    † : These headings are present only if the option "Only scraping?" is not selected.
    
    \caption{The figure shows some of the key headers in the CSV files. The posts are stored in a single file called "data.csv", with "Reddit post ID" as the primary key. Metadata retrieved from Spotify URL is stored in a separate CSV named after the Spotify URI (Unique Resource Identifier). The spotify URLs act as a foreign key referring to track metadata for URLs present in the Reddit posts.}
    \label{fig:csv_headers}
\end{figure}

After processing all of the data through these methods, the tool generates a CSV file which is automatically saved in the Muse-it folder. As shown in Figure \ref{fig:download_screen}, it contains all of the Reddit content collected like titles, posts, comments, time of creation and number of comments, along with generated metadata like topics, sentiments and emotions tagged with each post. The size of this CSV file depends on the length of posts, number of links and Spotify tracks found, but is expected approximately to be in the order of tens of megabytes. The track metadata of Spotify URLs are saved in separate CSV files. This allows the researcher to run any secondary experiments on the dataset of Spotify tracks and playlists mentioned in the retrieved data. Figure \ref{fig:csv_headers} shows the structure of the CSV files.

To avoid repeated searches on the same URL, the metadata retrieved for the tracks is stored in a common folder called "Spotify metadata", with separate folders for tracks, albums, and playlists. Metadata for each Spotify URL is stored in a separate CSV named after the corresponding Spotify URI (Unique Resource Identifier)\footnote{\url{https://developer.spotify.com/documentation/web-api/concepts/spotify-uris-ids}}.   

\section{Discussion}

Muse-it presents a robust solution for the extraction and analysis of music-related discourse on Reddit, transforming vast and unstructured social media data into datasets that are easily manageable for researchers. By automating the retrieval and organization of online discourse, the tool reduces the complexity that is traditionally associated with big data. It enables researchers to focus on extracting meaningful insights rather than grappling with data curation challenges. 

Muse-it's ability to organically combine Reddit discourse with rich track metadata from Spotify allows researchers to bridge the gap between qualitative insights and quantitative attributes. Muse-it can therefore serve as an invaluable tool for mixed-methods research in the field of musicology that empowers researchers to validate existing inferences in more ecologically valid settings and generate novel hypotheses. By lowering the technical barriers to big data analytics, Muse-it enables a broader community of researchers to explore the nuanced relationships between online discourse and music. It is also worth noting that the tool can be used by researchers outside the field of musicology for data retrieval and analysis.

The primary limitations of the tool lie in the restrictions present in the Reddit API. Reddit API limits extraction to 1000 posts per subreddit. Another limitation of Muse-it's approach is that it relies on Reddit users to explicitly have links to music content in their posts and comments. While Muse-it provides the discourse in textual format, a possible improvement in future versions would be the incorporation of text-based music identification techniques that can infer musical references even when direct links are absent. However, the integration of track metadata extraction from URLs in discourse remains a novel feature of Muse-it that bridges social media discussions with actual listening behaviors.

Reddit, as a platform, might contain a population skewed towards certain demographics and opinions. Despite having a language bias towards English, Reddit has many diverse linguistic communities. A query written in German, for example "traurige Musik", which is German for "sad music", would result in the retrieval of German subreddits. This allows researchers to use appropriate queries to overcome limitations like language when using Muse-it.

In the future, we plan to expand the track metadata extraction to include YouTube URLs. However, one of the primary challenges is to separate music-related URLs from non-music ones. Moreover, there can be non-music videos that have some music playing as a background track. Another feature we plan to include is the option to choose among different models for performing the classification tasks. As more researchers use our tool, we will utilize any feedback received to improve Muse-it further. 

\section{Ethics Statement}

Muse-it strictly adheres to the terms of services of Reddit API and follows their guidelines for data usage\footnote{\url{https://redditinc.com/policies/data-api-terms}}\footnote{\url{https://redditinc.com/policies/developer-terms}}. Throughout its pipeline, it upholds the anonymity of Reddit users, and does not access or utilize any personally identifiable information. It uses SpotDL only for the retrieval of textual metadata and does not access any copyright content. 

\bibliography{ISMIRtemplate}

\begin{thebibliography}{10}
\providecommand{\url}[1]{#1}
\csname url@samestyle\endcsname
\providecommand{\newblock}{\relax}
\providecommand{\bibinfo}[2]{#2}
\providecommand{\BIBentrySTDinterwordspacing}{\spaceskip=0pt\relax}
\providecommand{\BIBentryALTinterwordstretchfactor}{4}
\providecommand{\BIBentryALTinterwordspacing}{\spaceskip=\fontdimen2\font plus
\BIBentryALTinterwordstretchfactor\fontdimen3\font minus
  \fontdimen4\font\relax}
\providecommand{\BIBforeignlanguage}[2]{{%
\expandafter\ifx\csname l@#1\endcsname\relax
\typeout{** WARNING: IEEEtran.bst: No hyphenation pattern has been}%
\typeout{** loaded for the language `#1'. Using the pattern for}%
\typeout{** the default language instead.}%
\else
\language=\csname l@#1\endcsname
\fi
#2}}
\providecommand{\BIBdecl}{\relax}
\BIBdecl

\bibitem{if_i_like_2024}
T.~B.~M. Cao and T.~Bogers, ````if i like blank, what else will i like?'':
  Analyzing a human recommendation community on reddit,'' in \emph{Wisdom,
  Well-Being, Win-Win}, I.~Sserwanga, H.~Joho, J.~Ma, P.~Hansen, D.~Wu,
  M.~Koizumi, and A.~J. Gilliland, Eds.\hskip 1em plus 0.5em minus 0.4em\relax
  Cham: Springer Nature Switzerland, 2024, pp. 70--83.

\bibitem{asmr_article}
A.~Kovacevich and D.~Huron, ``Two studies of autonomous sensory meridian
  response (asmr): The relationship between asmr and music-induced frisson,''
  \emph{Empirical Musicology Review}, vol.~13, p.~39, 01 2019.

\bibitem{huron_2006}
\BIBentryALTinterwordspacing
D.~Huron, \emph{Sweet Anticipation: Music and the Psychology of
  Expectation}.\hskip 1em plus 0.5em minus 0.4em\relax The MIT Press, 04 2006.
  [Online]. Available: \url{https://doi.org/10.7551/mitpress/6575.001.0001}
\BIBentrySTDinterwordspacing

\bibitem{bhavyajeet}
\BIBentryALTinterwordspacing
B.~Singh, K.~Vaswani, S.~Paruchuri, S.~Saarikallio, P.~Kumaraguru, and
  V.~Alluri, ``“help! i need some music!”: Analysing music discourse \&
  depression on reddit,'' \emph{PLOS ONE}, vol.~18, no.~7, pp. 1--16, 07 2023.
  [Online]. Available: \url{https://doi.org/10.1371/journal.pone.0287975}
\BIBentrySTDinterwordspacing

\bibitem{sharon2024}
S.~Varghese, E.~Carlson, and V.~Alluri, ``Listening to the spectrum: Exploring
  music preferences and associations in autism spectrum disorder communities on
  reddit,'' 07 2024.

\bibitem{moisio2022just}
L.~Moisio, ``Just me and my music--the identity construction of christian metal
  fans on the subreddit r/christianmetal,'' 2022.

\bibitem{mishra-etal-2021-metal}
\BIBentryALTinterwordspacing
V.~Mishra, K.~Liew, E.~V. Epure, R.~Hennequin, and E.~Aramaki, ``Are metal fans
  angrier than jazz fans? a genre-wise exploration of the emotional language of
  music listeners on {R}eddit,'' in \emph{Proceedings of the 2nd Workshop on
  NLP for Music and Spoken Audio (NLP4MusA)}, S.~Oramas, E.~Epure,
  L.~Espinosa-Anke, R.~Jones, M.~Quadrana, M.~Sordo, and K.~Watanabe,
  Eds.\hskip 1em plus 0.5em minus 0.4em\relax Online: Association for
  Computational Linguistics, Nov. 2021, pp. 32--36. [Online]. Available:
  \url{https://aclanthology.org/2021.nlp4musa-1.7/}
\BIBentrySTDinterwordspacing

\bibitem{Veselovsky_Waller_Anderson_2021}
\BIBentryALTinterwordspacing
V.~Veselovsky, I.~Waller, and A.~Anderson, ``Imagine all the people:
  Characterizing social music sharing on reddit,'' \emph{Proceedings of the
  International AAAI Conference on Web and Social Media}, vol.~15, no.~1, pp.
  739--750, May 2021. [Online]. Available:
  \url{https://ojs.aaai.org/index.php/ICWSM/article/view/18099}
\BIBentrySTDinterwordspacing

\bibitem{camacho-collados-etal-2022-tweetnlp}
J.~Camacho-Collados, K.~Rezaee, T.~Riahi, A.~Ushio, D.~Loureiro, D.~Antypas,
  J.~Boisson, L.~Espinosa-Anke, F.~Liu, E.~Mart{\'\i}nez-C{\'a}mara
  \emph{et~al.}, ``{T}weet{NLP}: {C}utting-{E}dge {N}atural {L}anguage
  {P}rocessing for {S}ocial {M}edia,'' in \emph{Proceedings of the 2022
  Conference on Empirical Methods in Natural Language Processing: System
  Demonstrations}.\hskip 1em plus 0.5em minus 0.4em\relax Abu Dhabi, U.A.E.:
  Association for Computational Linguistics, Nov. 2022.

\bibitem{antypas2023robust}
D.~Antypas and J.~Camacho-Collados, ``Robust hate speech detection in social
  media: A cross-dataset empirical evaluation,'' 2023.

\bibitem{grootendorst2022bertopic}
M.~Grootendorst, ``Bertopic: Neural topic modeling with a class-based tf-idf
  procedure,'' \emph{arXiv preprint arXiv:2203.05794}, 2022.

\bibitem{reimers-2020-multilingual-sentence-bert}
\BIBentryALTinterwordspacing
N.~Reimers and I.~Gurevych, ``Making monolingual sentence embeddings
  multilingual using knowledge distillation,'' in \emph{Proceedings of the 2020
  Conference on Empirical Methods in Natural Language Processing}.\hskip 1em
  plus 0.5em minus 0.4em\relax Association for Computational Linguistics, 11
  2020. [Online]. Available: \url{https://arxiv.org/abs/2004.09813}
\BIBentrySTDinterwordspacing

\end{thebibliography}

\end{document}